\documentclass[aps, twocolumn, showpacs]{revtex4}
\begin{document}
\title{Viscous Spacetime Fluid and Higher Curvature Gravity}
\author { S. C. Tiwari\\
Institute of Natural Philosophy\\
c/o 1 Kusum Kutir Mahamanapuri,Varanasi 221005, India}
\begin{abstract}
 The Einstein field equation as an equation 
of state of a thermodynamical system of spacetime is reconsidered in the present Letter. We 
argue that a consistent interpretation leads us to identify scalar curvature and cosmological constant terms representing the bulk viscosity of the spacetime fluid. Since
Einstein equation itself corresponds to a near-equilibrium state in this interpretation invoking 
$f(R)$ gravity for nonequilibrium thermodynamics is not required. A logically consistent
generalization to include the effect of so called 'tidal forces' due to the Riemann curvature
is presented. A new equation of state for higher curvature gravity is derived and its
physical interpretation is discussed.
\end{abstract}
\maketitle

Higher curvature Lagrangians have been discussed as logically possible constructions
for gravitation since long \cite{1} , however current interest in such theories is
inspired by various shades of quantum gravity/effective field theory \cite{2}. Amongst
them $f(R)$ gravity, where $f$ is a nonlinear function of the scalar curvature $R$, is
considered as an attractive model to explain the observed cosmic acceleration
phase \cite{3}. The conflict of the metric formulation of $f(R)$ gravity with the
solar system observations has been claimed to get resolved in the Palatini formalism
\cite{4}, however it has been argued that this theory results in the appreciable
deviations from the microphysics (e.g. electron-electron scattering experiments), and
violation of the equivalence principle \cite{5}. It may be asked: Is there an
alternative approach to higher curvature gravity?

Departing from the variational principle approach, recently a thermodynamical derivation
of the $f(R)$ gravity has been proposed \cite{6} generalizing the previous derivation
of the Einstein field equation \cite{7}. Thermodynamics of spacetime in \cite{7} is
motivated by the black hole thermodynamics: it is assumed that similar to the black hole
entropy formula a universal entropy density $\alpha$ per unit horizon area for all local
Rindler horizons could be defined, and the Clausius relation $T dS=\delta Q$ of equilibrium
thermodynamics holds for all local horizons. Here the heat flow across the horizon
$\delta Q$ is defined to be the boost energy of the matter and $T$ is the Unruh
temperature. Jacobson then argues that since entropy is interpreted as horizon area
the Clausius relation would be satisfied provided the spacetime curvature in the 
presence of matter is such that the Einstein field equation holds. In \cite{6} this
approach is applied to the assumed entropy density proportional to $f(R)$, and it
is found that local energy conservation entails entropy production term, and hence
nonequilibrium thermodynamics of spacetime. Further the higher curvature equation of state
obtained is shown to be identical with the field equation derived from the $f(R)$
Lagrangian. 

Obviously rederivation of the known field equations by itself is not
of much significance unless new insights are gained. Unfortunately the paradigm
of field theory has overshadowed this promising thermodynamic approach in \cite{6,7}.
Moreover we have shown \cite{8} that covariant divergence law for the matter stress
tensor is satisfied in a conformally related spacetime, $\tilde{g}_{\mu\nu} =f(R) g_{\mu\nu}$
with no need to postulate bulk viscosity production term as is done by Eling et al
in \cite{6}, and therefore the issue of nonequilibrium thermodynamics becomes trivial.
Note that this result is independent of the unimodular perspective suggested in \cite{8}.

In this Letter spacetime as a thermodynamical system in the spirit of Jacobson's
approach is considered with the two-fold aim: 1) to gain new insights into the nature
of spacetime assumed to be some kind of fluid at large scale, and 2) to motivate
nonequilibrium thermodynamics in the case considered in \cite{7} and generalize it to include  
the effect of the so called 'tidal force'
caused by the Riemann curvature of spacetime \cite{9} and derive a new higher curvature 
equation of state.

First we argue that a consistent equation of state interpretation of the Einstein
field equation demands a careful reanalysis of the local equilibrium condition: the
$\Phi$ term in Eq.(6) below while easily interpreted in the field theoretic setting,
seems to require a different meaning since local energy conservation is assumed to be
a form of first law of thermodynamics \cite{6}. This term in the Einstein field equation 
in the absence of matter stress tensor is suggested to represent a dissipative
contribution to the stress tensor of vacuum. Analogous to the bulk viscosity
for the dissipative fluid \cite{10}, $\Phi$ term, therefore signifies viscous spacetime,
and the Einstein equation already corresponds to a quasi-equilibrium system.

Next we recall Bondi's argument \cite{11} noted in \cite{8} that the essence of 
observable gravitation is that relative acceleration varies at each spacetime point.
It would mean that even the assumption of local thermodynamic equilibrium for
local Rindler horizons would, in principle, fail. However the gravitation is so weak
that the correction due to Riemann curvature is proposed to be incorporated retaining
key elements of the Jacobson's approach in the following. The notion of local Rindler
horizon is used to define heat flow and entropy change in \cite{7}. The equivalence
principle enables locally flat spacetime at each spacetime point, and for an 
infinitesimal 2-surface element through the point one can assume vanishing shear
and expansion to the past (or inside) of the plane. The past horizon, the local 
Rindler horizon is assumed instantaneously stationary. This idea helps in
introducing an approximate local boost Killing vector field which can be related with the
horizon tangent vector and affine parameter along the null geodesic. Now instead
of employing the Jacobi deviation equation for relative acceleration, we assume that
the imprint of the tidal force is carried by the congruence of the null geodesics
via a correction term dependent on the Riemann curvature. Both nonuniform
temperature and viscosity lead to dissipation; since we do not know how to define
these quantities here instead of the entropy change we calculate change in the
thermodynamic potential $F=S T$. Heat flow and change in the horizon area are
calculated for local Rindler horizons to lowest order in the affine parameter as
is done in \cite{7}. A correction term proportional to the Riemann curvature
is introduced in the change $\delta F$. Equating $\delta Q$ and $\delta F$, and
imposing covariant divergence law for matter stress tensor finally give the desired
equation of state. The Letter concludes with a brief discussion on the physical
interpretation and prospects of this equation.

Briefly the main ideas of \cite{7} are as follows. The black hole formula, namely the proportionality between entropy
and the horizon area, is assumed to hold for all local Rindler horizons at each spacetime
point of the manifold M. Causal horizon at a point p is specified by a space-like
2-surface B, and the boundary of the past of B comprises of the congruences of null
geodesics. Assuming vanishing shear and expansion at p the past horizon of B is called
local Rindler horizon. The energy
flux across the horizon is used for heat energy, and calculated in terms of the boost energy of matter: define an approximate boost Killing vector field $\chi^\mu$ future pointing on the causal 
horizon, and related with the horizon tangent vector $k^\mu$ and affine 
parameter $\lambda$ by $\chi^\mu=-a \lambda k^\mu$. The heat flux to the past of B is given by
\begin{equation}
\delta Q=\int T_{\mu\nu} \chi^\mu d\Sigma^\nu
\end{equation}
The integral is taken over a small region of pencil of generators of the inside past horizon terminating at p.  If area element is $dA$ then $d\Sigma ^\nu=k^\nu d\lambda dA$, and Eq.(1)
becomes
\begin{equation}
\delta Q=-a \int T_{\mu\nu} k^\mu k^\nu \lambda d\lambda dA
\end{equation}
Change in the horizon area is given in terms of the expansion of the congruence of null geodesics generating the horizon $\delta A=\int \theta d\lambda dA$. The expansion of the null geodesics generating the horizon is given by the Raychaudhuri equation
\begin{equation}
\frac {d\theta}{d\lambda}=-\frac{\theta^2}{2}-\sigma_{\mu\nu}\sigma^{\mu\nu}-R_{\mu\nu}k^\mu k^\nu
\end{equation}
Assuming vanishing shear and neglecting $\theta ^2$ term we get the solution
\begin{equation}
\theta=-\lambda R_{\mu\nu} k^\mu k^\nu
\end{equation}
Assuming universal entropy density $\alpha$ per unit horizon area it is straightforward to
calculate the entropy change
\begin{equation}
\delta S=-\alpha \int R_{\mu\nu} k^\mu k^\nu \lambda d\lambda dA
\end{equation}
The condition that the Clausius relation is satisfied for all null vectors $k^\mu$ ,
and making use of the Unruh temperature $\hbar a/2\pi$ gives
\begin{equation}
R_{\mu\nu} + \Phi g_{\mu\nu}=(2\pi /\hbar\alpha) T_{\mu\nu}
\end{equation}
The unknown function $\Phi$ is determined using the covariant divergence law for the stress tensor and contracted Bianchi identity; Einstein equation with a cosmological constant (CC) $\Lambda$ is obtained. Here Newton's gravitational constant is identified as $G=1/4\hbar\alpha$, and
the function $\Phi$ is obtained to be
\begin{equation}
\Phi =-\frac{R}{2} +\Lambda
\end{equation}

In the light of thermodynamic derivation a logically consistent physical interpretation
of the equation of state is proposed here. In the Clausius relation integrands of
(2) and (5) have been equated for all null vectors to obtain Eq.(6). Since $T_{\mu\nu}$
is the matter stress tensor, from the integral form itself it is logical to infer that 
$R_{\mu\nu}$ is proportional to the stress of assumed spacetime fluid.
What does $\Phi$ term represent? In the derivation $\Phi$ term is needed to satisfy
the first law of thermodynamics (in the form of local energy conservation), and hence it
should correspond to an additional stress tensor for a dissipative process in Eq.(6).
Noting its formal similarity with the bulk viscosity stress tensor for an isotropic
fluid proportional to the tensor $\delta _{ij}$, see Eq.(8.4.42) in \cite{10} it
seems reasonable to identify $\Phi$ term in Eq.(6) as bulk viscosity tensor of the
spacetime. Deeper insight is gained analysing the CC term. Setting matter stress tensor 
zero and assuming a vanishing CC the Ricci flat spacetime becomes indistinguishable 
from $G_{\mu\nu}=R_{\mu\nu}-\frac{R}{2} g_{\mu\nu}=0$
as $R=0$. Thus vacuum Einstein equation with vanishing CC could be interpreted as an equation
of state of a perfect fluid. The presence of matter is suggested to cause a dissipative
process leading to a nonvanishing scalar curvature, i.e. a bulk viscosity to the
spacetime; geometrically scalar curvature arises via the contracted Bianchi identity 
so that the Einstein tensor satisfies the covariant divergence law. Here $R$ has
been given a thermodynamic significance.

What is the physical significance of $\Lambda$ ? In the light of its appearance in 
combination with $R$ in (7) this would correspond to the bulk viscosity treating
the whole $\Phi$ term as the bulk viscosity stress tensor of the spacetime. Thus
empty spacetime with nonvanishing CC would be like a viscous spacetime fluid. In 
Jacobson's approach the thermal behavior of quantum vacuum in flat spacetime is an
important ingredient. The time translation symmetry in Rindler wedge corresponds to
the boost symmetry of the Minkowskian spacetime; quantum vacuum in Minkowskian
spacetime as observed in the uniformly accelerating frame acquires thermal properties
of a thermal bath of real particles with the Unruh temperature \cite{12}. Assuming that 
quantum vacuum fluctuations would be significant only at Planck length scales a
universal entropy per unit horizon area and Unruh temperature have been assumed
in \cite{7}. This is a reasonable assumption, however in the light of thermodynamic
derivation it may be asked: Does there exist averaged out observable effect of
vacuum fluctuations? There is an interesting phenomenon in quantum optics \cite{13} :
atomic motion in light radiation in the so called optical molasses bears resemblance
with that of a particle in a viscous fluid. Analogous to this, envisaging gravitational
molasses having origin in quantum fluctuations in the spacetime bulk viscosity
associated with CC could be attributed to this effect. Another possibility is,
following suggestions in the literature, to treat CC as the energy density of
quantum vacuum fluctuations, $\Lambda = 8 \pi G {V_0} /{c^4}$ where $V_0$
is the vacuum expectation value. In this case, nonzero CC would cause dissipative
process similar to matter energy and constant scalar curvature would determine the
bulk viscosity of spacetime. We emphasize that both CC and quantum vacuum possess
rather speculative character in general relativity, and the microscopic nature
of the spacetime is unknown to us, hence the preceding discussion though quite
plausible, also remains speculative.

The question of bulk viscosity for the Einstein equation is also discussed in \cite{6};
the expected bulk viscosity of $3\hbar \alpha/4\pi$ is obtained from the $f(R)$ equation
in the Einstein frame. Authors argue it to be incorrect, and note the ambiguity of
sign. However negative bulk viscosity could be related with the acausal
teleological boundary conditions for the black hole \cite{14}. The stretched-horizon
formalism developed in \cite{14} gives useful insights on black hole physics; in 
particular, energy and momentum conservation laws for the membrane resemble with those of 
viscous fluid. The present interpretation that the Einstein equation represents a viscous
spacetime fluid could be viewed as a generalization, albeit a radical one, of the
membrane paradigm of \cite{14}.

Thermodynamic derivation of the Einstein equation based on arbitrary spacelike 2-surfaces
has been recently established \cite{15}. It is, therefore, reasonable to assert that
the preceding physical interpretation is of general validity. Clearly  viscosity
term implies that the system is not in thermodynamic equilibrium; moreover higher
curvature, e.g. $f(R)$ gravity, is not needed to motivate nonequilibrium thermodynamics.
The present interpretation however does offer an effective procedure to include the
effect of Riemann curvature as a correction to viscous stress tensor. Returning to the 
fluid analogy, in general the dissipative coupling coefficient $\eta_{ijkl}$ is a 
fourth rank viscosity tensor \cite{10}, and has symmetry of indices similar to
curvature tensor $R_{\mu\nu\alpha\beta}$. Assuming that the correction due to the
Riemann curvature is proportional to $R_{\mu\sigma\nu\rho}$ the product with the Ricci
tensor keeping in mind the symmetry of the indices would be 
$R^{\sigma\rho} R_{\mu\sigma\nu\rho}$. Thus we construct the simplest form of $\delta F$ 
\begin{equation}
\delta F=-\frac {\alpha\hbar a}{2\pi} \int [R_{\mu\nu}+\beta(R^{\sigma\rho} R_{\mu\sigma\nu\rho}
-\frac {R}{2} R_{\mu\nu})] k^\mu k^\nu \lambda d\lambda dA
\end{equation}
The last term in the square brackett in the integrand corresponds to the viscosity as identified above in Eq.(7) for the scalar curvature. This also makes
the application of covariant divergence law unambiguous. Using Clausius relation
and equating the integrands of (8) and (2) for all null vectors we get 
\begin{equation}
R_{\mu\nu}+\beta(R^{\sigma\rho} R_{\mu\sigma\nu\rho}-\frac {R}{2} R_{\mu\nu}) +\Psi g_{\mu\nu}
=\frac {2\pi}{\hbar \alpha} T_{\mu\nu}
\end{equation}
The unknown function $\Psi$ is determined taking the covariant divergence of (9) and
using the Bianchi identity and vanishing divergence of matter stress tensor. Use is
made of following relations
\begin{equation}
(R_{\mu\sigma\nu\rho} R^{\sigma\rho})^{: \mu}=(\frac {1}{4} R_{\sigma\rho}R^{\sigma\rho}
-\frac {1}{2} R_{:\alpha}^{~:\alpha})_{,\nu} +\frac {1}{2}(R_{;\mu\nu})^{:\mu}
\end{equation}

\begin{equation}
(RR_{\mu\nu})^{:\mu} = (R_{;\mu\nu})^{:\mu} + (\frac {R^2}{4} - R_{:\alpha}^{~:\alpha})_{,\nu}
\end{equation}
The expression for $\Psi$ is finally obtained to be
\begin{equation}
\Psi = -\frac {R}{2} + \frac {\beta R^2}{8} - \frac {\beta R^{\sigma\rho} R_{\sigma\rho}}{4}
\end{equation}
Substituting $\Psi$ in (9) we get the desired higher curvature equation of state
\begin{equation}
G_{\mu\nu} - \frac {\beta R}{2} (R_{\mu\nu} - \frac {R}{4} g_{\mu\nu}) +\beta R^{\sigma\rho}
(R_{\mu\sigma\nu\rho} - \frac {R_{\sigma\rho}}{4} g_{\mu\nu})= \frac {2\pi}{\hbar \alpha} T_{\mu\nu}
\end{equation}
Perusal of the higher curvature field equations in the literature \cite{2,3,4} shows that
Eq.(13) is a new result. This equation has some remarkable properties: 1) The arbitrary
constant $\beta$ could be adjusted to keep intact the established physical tests of
the Einstein equation since the higher curvature effect appears as a correction to
$G_{\mu\nu}$. 2) The trace of (13) leads to the same result as that of pure general
relativity. Note also that the first two terms are same as the ones in $R^2$ gravity
in the Palatini formalism which would make it to be in agreement with some of the results 
of that theory. However the third term is a new addition. And 3) In contrast
to the higher curvature field equations obtained from the action principle which are fourth order derivative equations, Eq.(13) is a second order derivative equation similar to that one
obtains in the Palatini version where the action is varied taking metric and affine connection 
as independent variables. These characteristics strongly suggest viability of Eq.(13) as higher
curvature gravity theory.

It is well known that the question of gravitational energy in general relativity
is quite complex \cite{16}, and even more so in higher curvature gravities \cite{2,17}.
An interesting consequence of (13) is that in the empty spacetime it reduces to
\begin{equation}
R_{\mu\nu} + \beta R^{\sigma\rho}(R_{\mu\sigma\nu\rho} - \frac {R_{\sigma\rho}}{4} g_{\mu\nu})
= 0
\end{equation}
What is the physical significance of the second term in Eq.(14)? In analogy with the
traceless stress tensor of electromagnetic field it is plausible to identify this
term to represent stress tensor of gravity radiation loss. Bondi presented a lucid
discussion \cite{18} on the inductive and wave transfer of gravitational energy in general
relativity, and considered axially-symmetric vacuum solution of Weyl and Levi-Civita for this purpose. Since the vacuum defined by Eq.(14) has different nature, it would be interesting to delineate inductive and wave energy transfers in this case. 

The problem of entropy production for $f(R)$ gravity has been extensively discussed in 
\cite{19}. What would be the entropy production in
the present case? There is another interesting question: Could one derive Eq.(13) in the
action formalism? We propose to investigate Palatini formalism for the action used by Deser
and Tekin \cite{2} elsewhere since the main theme of the present work is thermodynamic
approach to the spacetime.

In conclusion, we have argued that Einstein equation in thermodynamic approach
represents a viscous spacetime fluid, and derived a new equation of state for
higher curvature corrections.

I thank Prof. S. Deser for helpful correspondence, and Prof. Rong-Gen Cai  for drawing my attention to \cite{19}. Library facility of Banaras Hindu University is acknowledged.


\begin{thebibliography}{99}
\bibitem{1} A. S. Eddington, The Mathematical Theory of Relativity (C. U. P. 1924).
\bibitem{2} J. Ellis et al, Phys. Rev. D 59,103503(1999); S. Deser and B. Tekin, Phys.
Rev. Lett. 89,101101(2002).
\bibitem{3} S. M. Carroll et al, Phys. Rev. D 70,043528 (2004).
\bibitem{4} D. N. Vollick, Phys. Rev. D 68, 063510 (2003); X. Meng and P. Wang, Class.
Quantum Grav. 22, 23 (2005).
\bibitem{5} E. E. Flanagan, Phys. Rev. Lett. 92, 071101 (2004); G. J. Olmo, Phys.
Rev. Lett. 98, 061101 (2007).
\bibitem{6} C. Eling, R. Guedens, and T. Jacobson, Phys. Rev. Lett. 96,121301(2006)
\bibitem{7} T. Jacobson, Phys. Rev. Lett. 75,1260 (1995)
\bibitem{8} S. C. Tiwari, arXiv: gr-qc/0612099
\bibitem{9} S. W. Hawking and G. F. R. Ellis, The Large Scale Structure of Spacetime
(C.U.P. 1973).
\bibitem{10} P. M. Chaikin and T. C. Lubensky, Principles of Condensed Matter Physics
(C.U.P. 1995).
\bibitem{11} H. Bondi, Eu. J. Phys. 7,106 (1986)
\bibitem{12} W. G. Unruh, Phys. Rev. D 14, 870 (1976).
\bibitem{13} L. Mandel and E. Wolf, Optical Coherence and Quantum Optics 
(C.U.P. 1995) Sec. 15.8.1
\bibitem{14} R. H. Price and K. S. Thorne, Phys. Rev. D 33,915 (1986).
\bibitem{15} J. Makela and A. Peltola, arXiv: gr-qc/0612078
\bibitem{16} R. M. Wald, General Relativity (University of Chicago Press, 1984)
\bibitem{17} S. Deser and B. Tekin, arXiv: gr-qc/0701140
\bibitem{18} H. Bondi, Proc. R. Soc. London A 427, 249 (1990).
\bibitem{19} M. Akbar and Rong-Gen Cai, Phys. Lett. B 648, 243 (2007); arXiv: gr-qc/0612089

\end{thebibliography}
\end{document}